\documentclass[epj]{svjour}
\usepackage{graphics}
\begin{document}
\title{
Pionic Atom Spectroscopy in the ($d$,$^3$He) reaction at finite angles}
\author{
N.~Ikeno
\and H.~Nagahiro 
\and S.~Hirenzaki 
}
%
%
\institute{
Department of Physics, Nara Woman's University, Nara 630-8506, Japan
}
\date{Received: date / Revised version: date}
%
\abstract{
We study the formation of deeply bound pionic atoms in the ($d$,$^3$He)
reactions theoretically and show the energy spectra of the emitted
$^3$He at finite angles, which are expected to be observed
experimentally.
We find that the different combinations of the pion-bound 
and neutron-hole states dominate the spectra at different scattering 
angles because of the matching condition of the reaction. 
We conclude that the observation of the ($d$,$^3$He) reaction at finite
angles will provide the systematic information of the pionic 
bound states in each nucleus and will help to develop the study of the 
pion properties and the partial
restoration of chiral symmetry in nuclei.
} 
\maketitle
\section{Introduction}
\label{intro}
Deeply bound pionic atoms were discovered in the ($d$,$^3$He) spectra in 
1996~\cite{Yamazaki,Gilg,Itahashi}
following the theoretical predictions \cite{nd,d3He}.  
They have been considered to be one of the 
good systems to deduce the pion properties at finite density and to
obtain precise information on 
the partial restoration of chiral symmetry in nuclei
\cite{KSuzuki,Kolo,Jido}.

So far, the ($d$,$^3$He) spectra have been obtained in near recoilless
kinematics to observe 
the peaks of the pionic state formation with a neutron hole in the
quasi - substitutional configurations~\cite{Itahashi,KSuzuki,Geissel}.  
The Pb and Sn isotopes were used as the target nuclei in the 
experiments~\cite{Itahashi,KSuzuki,Geissel} 
since the spectra were expected to show the peak structures due to 
one dominant 
$[\pi \otimes n^{-1}]$ configuration based on 
the theoretical evaluations~\cite{Umemoto,Umemoto2}.  
The simple structure of the observed peak
is important and effective to deduce the pion binding energies 
precisely from the observed spectra.  
Actually, in Ref.~\cite{KSuzuki}, 
the binding energy and width of pionic atoms in Sn
isotopes were determined precisely and the partial restoration of chiral
symmetry was concluded based on the data.

To develop the studies of pion properties and symmetry
restoration in nuclei further,
we think that we need to obtain improved information from
experiments. 
Actually,
the systematic errors coming from 
the uncertainties of the neutron distribution and the
absolute energy calibration were reported in Ref.~\cite{KSuzuki}. 
In addition, we need more systematic information on bound states 
for unique determination of the pion-nucleus interaction,
which is required to fix the potential strength related to 
chiral symmetry.
Recently, it was reported in Ref.~\cite{Ikeno} that the 
simultaneous observation of various pionic bound states such as 
1$s$ and 2$s$ in the same nucleus 
may be helpful to reduce these errors and to develop our studies.
Along with this line, in actual experiments, 
the pionic 2$s$ state observation with 
the 1$s$ state has been expected in new high precision experiment in
RIBF/RIKEN \cite{Ikeno,RIBF1,RIBF2}.

In this paper, we consider theoretically the new possibility to observe 
various pionic bound states in the same nuclei by observing the ($d$,$^3$He) 
spectra at finite angles together with the forward direction.  
These observations at finite angles are expected to be possible in 
the experiments at RIBF/RIKEN~\cite{Itahashi2}.  
As well-known in the nuclear inelastic scatterings,
we can expect to have the manifestation of different 
subcomponents of pion and neutron hole states at finite angles due to
the matching condition with different momentum transfer, 
and expect to determine the binding energies and widths of various
pionic states simultaneously in each nucleus.
The momentum transfer dependence of the effective number was studied in
Ref.~\cite{d3He} for different incident energies.
The angular dependence of the resonance peak structure
will be helpful to determine the angular momentum quantum numbers 
of the populated states directly from the ($d$,$^3$He) data.


\section{Formulation}
\label{formula}
We modify slightly the theoretical model used in
Refs.~\cite{d3He,Umemoto,Umemoto2} to study the 
angular dependence of the ($d$,$^3$He) spectra.
The modifications are implemented by including the kinematical
correction factors $K$ in Eq.~(\ref{cross}) as explained below.
A similar consideration was also given in Ref.~\cite{Koike} for 
the ($K^-, N$) reaction.

We start our discussion by considering the
$d + n \rightarrow {^3{\rm He}} + \pi^-$ reaction, 
which is the elementary pion production process for the
formation of the pionic atoms in the ($d$,$^3$He) reaction.
Our model is based on the relativistic cross section
formula~\cite{Bjorken} with the Lorentz invariant 
pion production amplitude $T(p_d, p_n, p_{\rm He}, p_\pi)$.
The cross section can be written as,
\begin{eqnarray}
d \sigma &=& 
\frac{1}{v_{\rm rel}} \frac{M_d}{E_d} \frac{M_n}{E_n} 
|T(p_d, p_n, p_{\rm He}, p_\pi)|^2  \nonumber \\
&\times&
(2 \pi)^4 \delta^{4}(p_d + p_n - p_{\rm He} - p_\pi)  
\frac{M_{\rm He}}{E_{\rm He}} \frac{d {\vec p}_{\rm He}}{(2 \pi)^3}
\frac{1}{2 E_\pi} \frac{d {\vec p}_\pi}{(2 \pi)^3}, \nonumber \\
\label{dsigma}
\end{eqnarray}
where the subscript of the momentum $p$, the energy $E$ and 
the mass $M$ indicates each particle participating
in the two body reaction $d + n \rightarrow {^3{\rm He}} + \pi^-$.
The relative velocity $v_{\rm rel}$ between $d$ and $n$ can be written as,
\begin{equation}
 v_{\rm rel} =\frac{1}{2 E_d E_n} \lambda^{1/2}(s,M^2_d,M^2_n),
\end{equation}
where $s$ is the Mandelstam variable and
$\lambda$ is the K\"{a}llen function. 

The elementary cross section in the center of mass (CM) frame and
the laboratory frame at scattering angle
$\theta^{\rm lab}_{d {\rm He}}$, which is required for the 
effective number approach, are written as
\begin{eqnarray}
\left( \frac{d \sigma}{d \Omega_{\rm He}} \right)^{\rm CM}_{\rm ele} 
&=&  
\frac{|T(p_d, p_n, p_{\rm He}, p_\pi)|^2 M_d M_n M_{\rm He}}
{(2 \pi)^2 \lambda^{1/2}(s,M^2_d,M^2_n)} \nonumber\\
&\times&
\frac{2 |{\vec p}^{\rm CM}_{\rm He}|^2}{\lambda ^{1/2}(s,M^2_{\rm He},M^2_\pi)},
 \label{dsigma_CM}
\end{eqnarray}
and
\begin{eqnarray}
\left( \frac{d \sigma}{d \Omega_{\rm He}} \right)^{\rm lab}_{\rm ele} 
&=& 
\frac{|T(p_d, p_n, p_{\rm He}, p_\pi)|^2 M_d M_n M_{\rm He}}
{(2 \pi)^2 \lambda^{1/2}(s,M^2_d,M^2_n)}\nonumber\\
&\times&
\left[ \frac{|{\vec p}_{\rm He}|^ 2} 
{ E_\pi |{\vec p}_{\rm He}| + E_{\rm He}(|{\vec p}_{\rm He}|- |{\vec
p}_{d}| \cos\theta_{d{\rm He}})}  \right]^{\rm lab}, \nonumber\\
\label{dsigma_lab}
\end{eqnarray}
where the superscripts `lab' and `CM' indicate the frame where the
kinematical variables should be evaluated.
The ratio of these cross sections 
can be written as,
\begin{eqnarray}
&&\left ( \frac{d \sigma}{d \Omega_{\rm He}} \right)_{\rm ele}^{\rm lab} 
\big{/} \left( \frac{d \sigma}{d \Omega_{\rm He}} \right)_{\rm ele}^{\rm CM} 
\nonumber\\ 
&=& \left[ 
\frac{|{\vec p}_{\rm He}|^2}
{ E_\pi |{\vec p}_{\rm He}| + E_{\rm He} (|{\vec p}_{\rm He}|-|{\vec p}_d| 
\cos \theta_{d{\rm He}})} \right]^{\rm lab} \nonumber\\
&\times&
\frac{\lambda ^{1/2}(s,M^2_{\rm He},M^2_\pi)}{2 |{\vec p}^{\rm CM}_{\rm He}|^2}.
\label{dsigma_lab_CM}
\end{eqnarray}
This expression can be simplified as, 
\begin{equation}
\left( \frac{d \sigma}{d \Omega_{\rm He}} \right)_{\rm ele}^{\rm lab} 
\big{/} \left( \frac{d \sigma}{d \Omega_{\rm He}} \right)_{\rm ele}^{\rm CM} 
= \frac{|{\vec p}^{\rm lab}_{\rm He}|^2}{|{\vec p}^{\rm CM}_{\rm He}|^2},
\end{equation}
at forward angles $\theta_{d{\rm He}}=0^\circ$ 
as shown in Ref.~\cite{nd,d3He}. 
The experimental elementary cross sections in the CM frame 
were reported in Ref.~\cite{ele_data}.

Then, we consider the pionic bound state formation cross section for the
heavy nuclear target case in the same framework. 
In the effective number approach used in 
Refs. \cite{d3He,Umemoto,Umemoto2}, the elementary
cross section is factorized from the bound state formation cross section
as shown below.
The bound state formation cross section for the heavy nuclear target case
can be written in the laboratory frame as, 
\begin{eqnarray}
d \sigma =  \sum_{f} \frac{V^2}{v_{\rm rel}} 
\frac{1}{VT} |S_{fi}|^2 
\frac{V}{(2\pi)^3} d {\vec p_{\rm He}},
\label{dsig-A}
\end{eqnarray} 
with the S matrix, 
\begin{eqnarray}
S_{fi}&=& \int dt d{\vec x} 
 \sqrt{\frac{M_{\rm He}}{E_{\rm He}}} \frac{1}{\sqrt{V}} 
 e^{i E_{\rm He} t} \chi^{*}_{\rm He}(\vec{x})
 \sqrt{\frac{1}{2 E_\pi}} e^{i E_\pi t} \phi^{*}_\pi({\vec x})\nonumber\\
&\times& 
 i T(p_d, p_n, p_{\rm He}, p_\pi)\nonumber\\
&\times& 
 \sqrt{\frac{M_d}{E_d}} \frac{1}{\sqrt{V}} 
 e^{-i E_d t} \chi_{d}(\vec{x})
 \sqrt{\frac{M_n}{E_n}} e^{-i E_n t} \psi_n({\vec x}),
\label{S-nucl_1}
\end{eqnarray}
where $\phi_\pi$ and $\psi_n$ indicate the wavefunctions of the pion
bound state in the daughter nucleus and the neutron bound state in the
target nucleus, respectively.
The wavefunctions of the projectile ($d$) and the ejectile
($^3$He) are denoted by $\chi^{*}_{\rm He}$ and $\chi_d$.
$T(p_d, p_n, p_{\rm He}, p_\pi)$ indicates the same meson production
amplitude appeared in Eq.~(\ref{dsigma}) 
and describe the $d + n \rightarrow {^3{\rm He}} +\pi^{-}$ 
transition.
$V$ and $T$ indicate the spatial volume and the time interval for
the transition~\cite{Bjorken}.
We assume that only the neutron in the target nucleus participates in 
the reaction and other nucleons are spectators.
The formation cross section of the pion bound states in the laboratory
frame can be expressed as, 
\begin{eqnarray}
\left( \frac{d^{2}\sigma}{dE_{\rm He} d\Omega_{\rm He}} \right)^{\rm lab}_{A}
&=&
\sum_{f}  \biggl[ 
\frac{1 }{ 2 (2 \pi)^2 }
\frac{ M_d M_n M_{\rm He}}{ E_n E_\pi}
\frac{|{\vec p}_{\rm He}| }{|{\vec p}_d| }\nonumber\\
&\times&
|T(p_d, p_n, p_{\rm He}, p_\pi)|^2 
\frac{\Gamma}{2\pi}\frac{1}{\Delta E^{2}+{\Gamma}^{2}/4}\nonumber\\
&\times&
\left|\int  d{\vec x} 
 \chi^{*}_{\rm He}({\vec x}) \phi^{*}_\pi({\vec x})
 \psi_n ({\vec x}) \chi_d({\vec x}) \right|^2 
\biggl]^{\rm lab}_A ,\nonumber\\ 
\label{crossA}
\end{eqnarray}
where the subscript `$A$' indicates the ($d$,$^3$He) reaction for 
the nuclear target.
Here, we assume that 
the target nucleus is so heavy that we can ignore the recoil energy
of the nucleus and that we can consider
the CM and the laboratory frames to be the same.
The $\delta$-function responsible for the energy conservation is
replaced by the Lorentz distribution function 
$\displaystyle \frac{\Gamma}{2\pi}\frac{1}{\Delta E^{2}+{\Gamma}^{2}/4}$
to account for the width
$\Gamma$ of the pion bound state, where 
$\Delta E$ is defined as $\Delta E=E_{\rm He} +E_\pi - E_d -E_n$. 
$E_n$ is the energy of neutron in the target nucleus and defined as
$E_n=M_n-S_n (j_n)$ with the neutron separation energy $S_n$ from the
$j_n$ single particle level and the neutron mass $M_n$.
$E_\pi$ is the energy of the pion bound state defined as 
$E_\pi=M_\pi-B.E.(\ell_\pi)$
with the pion mass $M_\pi$ and the binding energy $B.E.$ of the bound
state indicated by $\ell_\pi$.
The reaction $Q$-value can be expressed as
$Q=\Delta E - M_\pi + B.E.(\ell_\pi)-S_n (j_n) +(M_n + M_d -M_{\rm He})$, 
where $M_n + M_d -M_{\rm He}=6.787$ MeV.

We express the effective number for each eigen state of the total
spin of neutron-hole and pion-particle states in the daughter nuclei.   
We decompose the sum of the final states as
\begin{equation}
\displaystyle \sum_{f} \rightarrow \sum_{ph} \sum_{J M},
\end{equation}
where $\displaystyle \sum_{ph}$ indicates the
sum of all combinations of the pion bound states and 
neutron hole states in the final states, 
and $\displaystyle \sum_{JM}$ indicates the sum of
the states with different total angular momentum for each
combination of bound pion and neutron hole state.
By neglecting the residual interaction effects for
the final state energies,
we can rewrite the bound state formation cross section as,
\begin{eqnarray} 
\left( \frac{d^{2}\sigma}{dE_{\rm He} d\Omega_{\rm He}} 
\right)^{\rm lab}_{A} 
&=&
\left (\frac{d\sigma}{d\Omega_{\rm He}}\right )^{\rm lab}
_{\rm ele} \nonumber\\
&\times&
\sum_{ph} K 
\frac{\Gamma}{2\pi}\frac{1}{\Delta E^{2}+{\Gamma}^{2}/4}
N_{\rm{eff}},
\label{cross}
\end{eqnarray}  
where we have used the elementary cross section 
 $\displaystyle \left( \frac{d\sigma}{d\Omega_{\rm He}} \right)
^{\rm lab}_{\rm ele}$
instead of $|T|^2$. 
The effective number $N_{\rm eff}$ defined as,
\begin{equation}
N_{\rm{eff}} =
\sum_{J M} 
\left| 
\int d{\vec r} \chi^{\ast}_{\rm He}({\vec r})
[\phi^{\ast}_{\ell_\pi}({\vec r})\otimes
\psi_{j_n}({\vec r})]_{JM}\chi_{d}({\vec r}) \right|^{2}.
\label{Neff}
\end{equation}
We introduce the distortion effects to the wavefunctions 
$\chi^{*}_{\rm He}$ and $\chi_d$ by Eikonal
approximation as described in Refs.~\cite{d3He,Umemoto,Umemoto2}.
The kinematical correction factor $K$ is derived by factorizing the
elementary cross section in the laboratory frame Eq.~(\ref{dsigma_lab})
from the bound state
formation cross section for the heavy nuclear target case Eq.~(\ref{crossA}), 
and is defined as,
\begin{eqnarray}
K=\left[ \frac{|{\vec p}^A_{\rm He}|}{|{\vec p}_{\rm He}|} 
\frac{E_n E_\pi} {E^A_n E^A_\pi} 
\left(1+\frac{E_{\rm He}}{E_\pi}\frac{|{\vec p}_{\rm He}|-|{\vec p}_d| 
{\rm cos}\theta_{d{\rm He}}}{|{\vec p}_{\rm He}|} \right) \right]^{\rm lab},
\nonumber\\
\label{K}
\end{eqnarray}
where $A$ indicates the momentum and energy which should be evaluated
in the kinematics of the nuclear target case.
Since we consider the heavy target limit, the scattering angle of
nuclear target case does not appear in Eq.~(\ref{K}).
It should be noted that in addition to the wavefunctions
$\phi_{\ell_\pi}$ and $\psi_{j_n}$,
$\Delta E$, $\Gamma$ and $K$ also depend on the final
`$ph$' combination through the binding energy and width of the pion
bound states and neutron separation energy.
The $K$ factor is evaluated by the energy and momentum at the kinematics
satisfying the condition $\Delta E=0$ for each `$ph$' combination.
This correction factor is $K=1$ for the recoilless kinematics at 
$\theta_{d{\rm He}}^{\rm lab}=0^\circ$ with $S_n =0$ and $B.E.=0$.

Here, we have assumed the same meson production amplitude 
$T(p_d, p_n, p_{\rm He}, p_\pi)$ for
the pion bound state formation process in the heavy nucleus 
as the elementary process by the impulse approximation.
We need to know the microscopic mechanisms of the meson production to
evaluate the possible modifications of 
the $T$ amplitude beyond the kinematical corrections.

As a summary of this section, we use Eq.~(\ref{cross}) to calculate 
the angular dependence of the pionic atom formation spectra.
This formula newly include the kinematical correction factor $K$ which
has a certain contribution to the angular dependence of the cross section
as we will see later.
As for the elementary cross section, we obtain 
$\displaystyle \left(\frac{d\sigma}{d\Omega_{\rm He}}\right)^{\rm lab}_{\rm ele}$ 
in Eq.~(\ref{cross}) by the formula Eq.~(\ref{dsigma_lab_CM}) from the
cross section in the CM frame which have been determined experimentally~\cite{ele_data}.


\section{Numerical Results and Discussions} 
\label{result}
First, we consider the angular dependence of the elementary cross
section. 
As the elementary cross section, we have used the experimental data of  
$p + d \rightarrow t + \pi^{+}$ reaction reported in 
Ref~\cite{ele_data} in the CM frame by isospin symmetry.
We parameterize the data shown in Fig.~5 in Ref~\cite{ele_data} as,  
\begin{equation}
\left( \frac{d \sigma}{d \Omega} \right)_{\rm ele}^{\rm CM}=
 10^{ -0.852 \hspace{1mm} \cos(\pi - \theta_{d {\rm He}}^{\rm CM})+ 0.051 } 
\hspace{3mm} [\mu {\rm b/sr}],
\label{ele_CM}
\end{equation}
and  transfer the cross section to that in the lab frame by 
Eq.~(\ref{dsigma_lab_CM}).
We show the values of elementary cross sections used in this paper in Table~\ref{tab:1}.

\begin{table}
\caption{Elementary cross sections in CM frame 
$(d \sigma/d \Omega)_{\rm ele}^{\rm CM}$ 
and those in lab frame $(d \sigma/d \Omega)_{\rm ele}^{\rm lab}$ 
in unit of [$\mu$b/sr].
The scattering angles $\theta_{d {\rm He}}^{\rm lab}$ and 
$\theta_{d {\rm He}}^{\rm CM}$ indicate
the angles between deuteron and $^3{\rm He}$ in the 
$d +n \rightarrow {^3{\rm He}} + \pi^{-}$ reactions in CM and lab frames, 
respectively.
We use Eq.~(\ref{ele_CM}) to fit the experimental data in Ref.~\cite{ele_data}.}
\label{tab:1}       
\begin{center}
\begin{tabular}{c||cccc}
\hline\noalign{\smallskip}
$\theta^{\rm lab}_{d {\rm He}}$ & $0.0^\circ$ & $1.0^\circ$  & $2.0^\circ$ & $3.0^\circ$  \\
$\displaystyle \left(\frac{d \sigma}{d \Omega} \right)_{\rm ele}^{\rm lab}$ & $1.77 \times 10^{3}$  & $1.70\times 10^{3}$  & $1.50\times 10^{3}$ & $1.22\times 10^{3}$ \\
\noalign{\smallskip}\hline\noalign{\smallskip}
$\theta^{\rm CM}_{d {\rm He}}$ & $0.0^\circ$ & $15.0^\circ$  & $31.0^\circ$ & $49.5^\circ$ \\
$\displaystyle \left(\frac{d \sigma}{d \Omega} \right)_{\rm ele}^{\rm CM}$  &  8.00  & 7.48   & 6.05  & 4.02  \\
\noalign{\smallskip}\hline
\end{tabular}
\end{center}
\end{table}

\begin{figure}
\begin{center}
\resizebox{0.42\textwidth}{!}{%
  \includegraphics{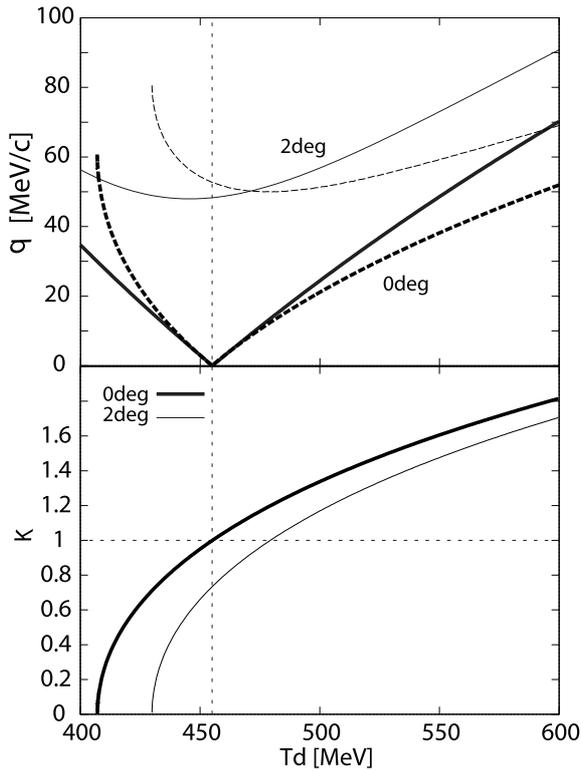}
}
\caption{
Momentum transfer ${\vec q}$ (upper) and kinematical correction
 factors $K$ (lower) of the ($d, ^3$He)
reaction at $\theta^{\rm lab}_{d {\rm He}}=0^\circ$ (thick lines) 
and $2^\circ$ (thin lines) 
plotted as functions of the incident deuteron kinetic energy $T_d$.
In the upper frame,
the solid lines show the results for the formation of $\pi$ meson  
in the heavy target nucleus,
and the dashed lines those for the elementary process.
The pion binding energy $B.E.$ and the neutron separation energy $S_n$ 
from the target nucleus are assumed to be 0 MeV.
The vertical line indicates the incident energy $T_d$ 
satisfying the recoilless condition at $\theta^{\rm lab}_{d {\rm He}}=0^\circ$.
}
\label{fig:1}       
\end{center}
\end{figure}

We, then, show the differences of the kinematics between 
the elementary process 
$d + n \rightarrow  {^3{\rm He}} + \pi^{-}$ 
and the meson bound state production
$A (d, {^3{\rm He}})(A-1) \otimes \pi^{-}$ with a heavy nucleus target.
We show in Fig.~\ref{fig:1} the momentum transfer $\vec{q}$ at
$\theta^{\rm lab}_{d{\rm He}}=0^\circ$ and $2^\circ$
for the
elementary process and the reaction with the heavy nuclear target for pion
production with the meson binding energy $B.E.=0$ and the neutron
separation energy $S_n =0$.
We find that the recoilless condition ($|{\vec q}|=0$) is only satisfied at
$\theta^{\rm lab}_{d{\rm He}}=0 ^\circ$ at $T_d = 455$ MeV for both reactions.
At different incident energy, the momentum transfer takes obviously
different value for these two reactions.
We also show the incident deuteron energy dependence of the $K$ factor
in Fig.~\ref{fig:1}.
We found that the $K$ factor is 1 for the recoilless kinematics with
$B.E.=0$ and $S_n =0$ as naturally expected.
The $K$ factor is $0$ at $T_d =407$ MeV and $T_d =430$ MeV for 
$\theta^{\rm lab}_{d {\rm He}}=0^\circ$ and $2^\circ$, respectively.
These energies correspond to the pion production threshold for the
elementary process.
And the $K$ factor increases with the incident deuteron kinetic energy $T_d$.

\begin{figure}
\begin{center}
\resizebox{0.45\textwidth}{!}{%
  \includegraphics{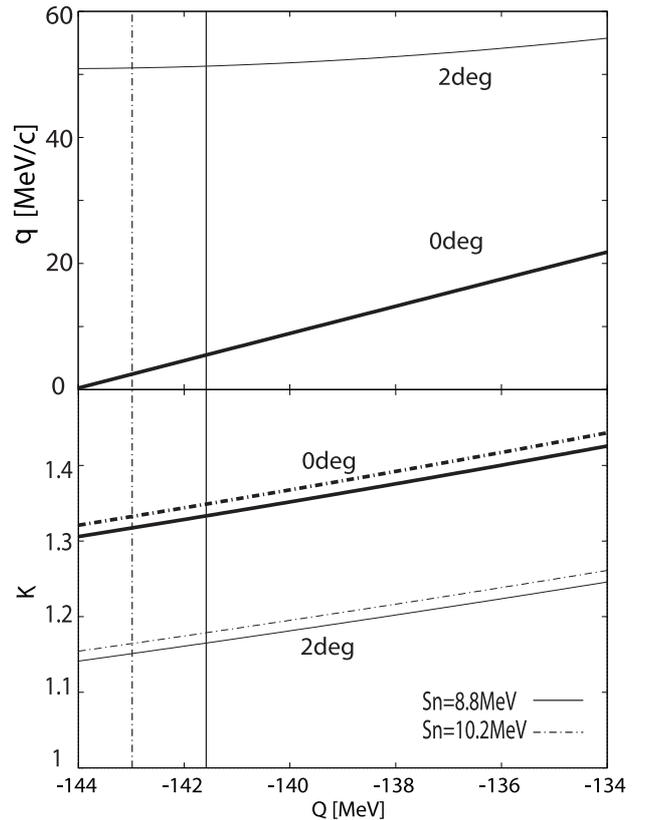}
}
\caption{
Momentum transfer ${\vec q}$ (upper) and kinematical correction
factors $K$ (lower) of the ($d, ^3$He) reaction 
at $\theta^{\rm lab}_{d {\rm He}}=0^\circ$ and $2^\circ$ 
for the formation of $\pi$ meson bound states in the heavy target nucleus   
plotted as functions of the reaction $Q$-value.
The incident deuteron kinetic energy is fixed to be $T_d = 500$ MeV.
The neutron separation energy $S_n$ is fixed to be 8.8 MeV (solid lines) and
10.2 MeV (dash-dotted lines).
The vertical lines indicate the $\pi$ meson production threshold for
$S_n =8.8$ and 10.2 MeV cases.
}
\label{fig:2}       
\end{center}
\end{figure}

In Fig.~\ref{fig:2}, 
we show the momentum transfer for the nuclear target case 
and the $K$ factor for $T_d =500$ MeV at 
$\theta^{\rm lab}_{d{\rm He}}=0^\circ$ and $2^\circ$ as functions 
of the reaction $Q$-value, which is the kinematics considered
in the missing mass spectroscopy of the deeply bound pionic atoms
in this article.
We find that the $K$ factor gradually
varies about 10\% within the $Q$-value range
considered here for both $\theta^{\rm lab}_{d{\rm He}}=0^\circ$ and 
$2^\circ$ cases, 
and the $K$ factor decreases about 20\% at 
$\theta^{\rm lab}_{d{\rm He}}=2^\circ$ from 
the value at $\theta^{\rm lab}_{d{\rm He}}=0^\circ$.
Thus, we think the $K$ factor should be introduced to study the angular
dependence of the ($d$,$^3$He) spectra.

\begin{figure}
\begin{center}
\resizebox{0.45\textwidth}{!}{%
  \includegraphics{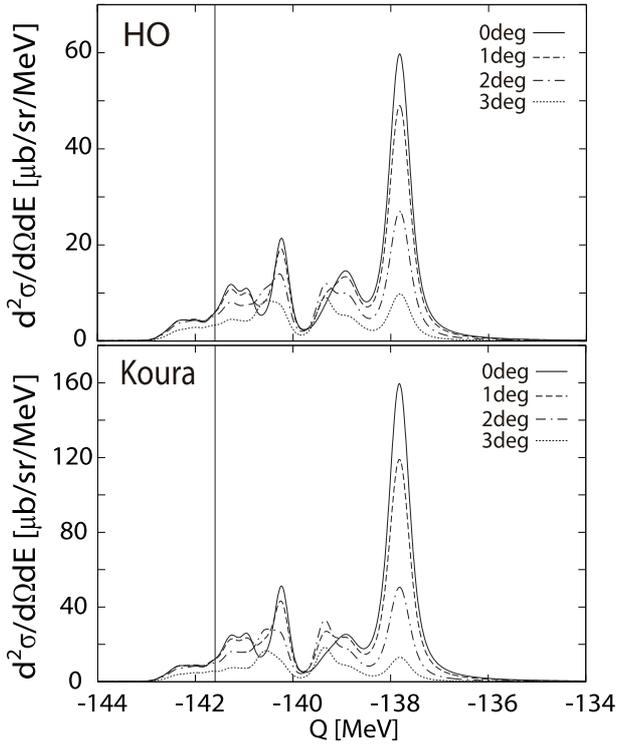}
}
\caption{
Calculated $^{122}$Sn($d, ^{3}$He) spectra 
for the formation of the pionic bound states 
at $\theta^{\rm lab}_{d{\rm He}}=0^\circ$ (solid lines), $1^\circ$(dash lines), 
$2^\circ$ (dash-dotted lines) and $3^\circ$ (dotted lines) 
plotted as functions of the reaction $Q$-value.
The harmonic oscillator (upper) and the phenomenological~\cite{koura} (lower)
neutron wavefunctions are used. 
The incident deuteron kinetic energy is fixed to be $T_d = 500$ MeV.
The instrumental energy resolution is assumed to be 300 keV FWHM.
The vertical line indicates the pion production threshold $Q=-141.6$ MeV.
}
\label{fig:3}       
\end{center}
\end{figure}

In Fig.~\ref{fig:3}, we show the calculated ($d, ^3{\rm He})$ spectra
at finite angles for the bound pionic states formation.
Two kinds of the neutron wavefunction have been used for the
calculations, which are the harmonic oscillator wavefunction and the
calculated wavefunction using the neutron potential reported in
Ref.~\cite{koura}.
Though the absolute value of the calculated cross sections depend on
the neutron wavefunction used as shown in Ref.~\cite{Ikeno}, the
angular dependence of the spectra resembles each other.
We find that the spectra have a strong angular dependence and the shape
of the spectra are much different at finite angles from that at
$0^\circ$.
The largest peak structure at $Q=-137.8$ MeV in the forward spectra is
strongly suppressed at finite angles and the spectra show the structure 
of three peaks at $\theta^{\rm lab}_{d{\rm He}} \geq 2^\circ$.
The overall strength of the spectra has also angular dependence and is
smaller for larger angles.

\begin{figure*}[h]
\begin{center}
\resizebox{0.95\textwidth}{!}{%
  \includegraphics{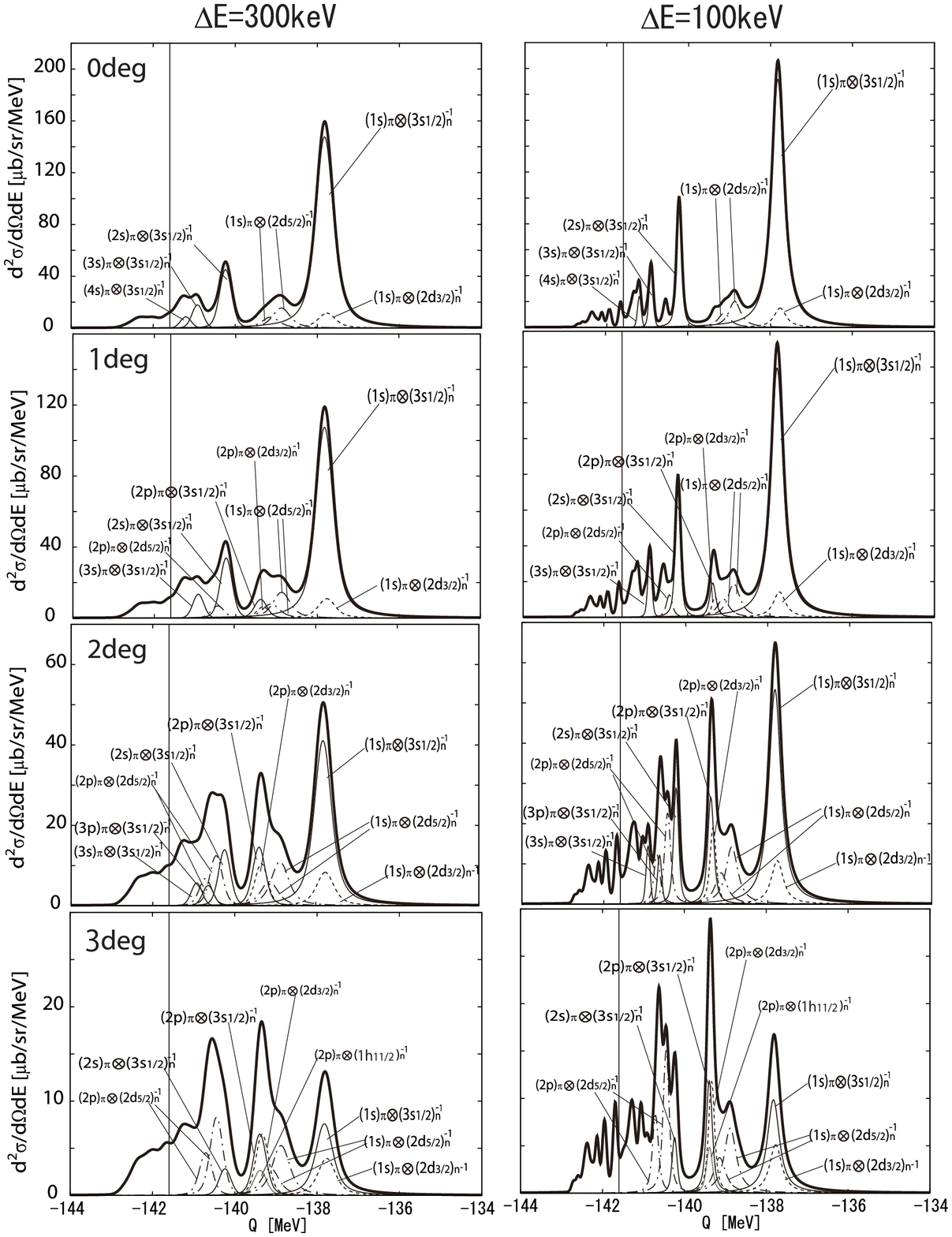}}
\caption{
Calculated $^{122}$Sn($d, ^{3}$He) spectra for the formation of the
pionic bound states
at $\theta_{d{\rm He}}^{\rm lab}=0^\circ$, $1^\circ$, $2^\circ$ and $3^\circ$  
are plotted as functions of the reaction $Q$-value.
Dominant subcomponents $[(n \ell)_{\pi} \otimes (n \ell_j)^{-1}_n]$
are indicated in the figure.
The phenomenological neutron wavefunctions~\cite{koura} are used. 
The instrumental energy resolution is assumed to be 300 keV FWHM 
(left) and 100 keV FWHM (right).
The vertical line indicates the pion production threshold $Q=-141.6$ MeV.}
\label{fig:4}       
\end{center}
\end{figure*}

In Fig.~\ref{fig:4}, we show the dominant subcomponents of ($d$,$^3{\rm He})$
spectra for each scattering angle $\theta^{\rm lab}_{d{\rm He}}$. 
Here, we have used the neutron wavefunction obtained by the potential
in Ref.~\cite{koura} and we have also shown the spectra with better
energy resolution 100 keV case. 
At $\theta^{\rm lab}_{d{\rm He}}=0^\circ$, since the reaction is close
to recoilless, the peaks of [$(1s)_\pi \otimes (3s_{1/2})^{-1}_n$] and
[$(2s)_\pi \otimes (3s_{1/2})^{-1}_n$] subcomponents appear clearly in
the spectra for both energy resolution cases. 
The contributions of $(3s)_\pi$ and $(4s)_\pi$ also show the isolated
peaks for better energy resolution (100 keV) case even though they are small. 
At $\theta^{\rm lab}_{d{\rm He}}=1^\circ$, the contribution for
[$(2s)_\pi \otimes (3s_{1/2})^{-1}_n$] is suppressed and can only be seen
clearly in a better energy resolution case. 
At larger angles, the pionic ($2p$) state contributions become
relatively larger
and dominate the peak structure around $Q=-139.4$ MeV and $-140.4$ MeV. 
We can expect to observe the peak structure composed from 
[$(2p)_\pi \otimes (3s_{1/2})^{-1}_n$],  [$(2p)_\pi \otimes (2d_{3/2})^{-1}_n$] 
and
[$(2p)_\pi \otimes (1h_{11/2})^{-1}_n$] subcomponents. 
Though, the separation energies of these 3 neutron levels differ from each
other only within 60 keV~\cite{Ikeno} and their contributions can not be
distinguished, we can expect to deduce the information on the pionic
2$p$ state.
These contributions of pionic 2$p$ state can not be seen in the spectra
at $\theta^{\rm lab}_{d{\rm He}}=0^\circ$ since they are hidden in the
tail of the large 1$s$ contributions.
At finite angles, due to the significant suppression of
the 1$s$ contributions, the 2$p$ contributions can be observed even if
they are a little smaller than those at the forward angle.
Thus, to observe the spectra at finite angle is valuable.

Finally, we comment on the small peak structures in the pionic unbound
energy region which appear in better energy resolution cases in
Fig.~\ref{fig:4} right panels.
They are the contributions of lightly bound pionic atom formation with 
one of the deep neutron hole states corresponding to the excited states of the
daughter nucleus such as 
$[(3p)_{\pi} \otimes (2d_{5/2})^{-1}_n]$,
$[(4p)_{\pi} \otimes (2d_{5/2})^{-1}_n]$,
$[(3d)_{\pi} \otimes (2d_{5/2})^{-1}_n]$,
$[(4d)_{\pi} \otimes (2d_{5/2})^{-1}_n]$ and so on.
The wavefunctions of the lightly bound pionic atom have large spatial
dimensions and small overlaps with nucleon wavefunction, and thus, both
widths and formation cross sections of these bound states are small.
Therefore, we can observe contributions of these states formations only in
the spectra with better energy resolution.

\section{Summary}
We study the formation of deeply bound pionic atoms in the ($d$,$^3$He)
reactions theoretically and show the angular dependence of the expected
spectra, which can be observed in experiments. 

We develop the formula to include the kinematical correction factors to
the effective number approach to obtain more realistic angular
dependence of the ($d$,$^3$He) spectra. 
We show the behavior of the kinematical correction factor and the angular
dependence of the $^{122}$Sn($d$,$^3$He) spectra at $T_d =500$ MeV for
the formation of the pionic atoms. 

We find that the spectra are dominated by the subcomponents including
$(2p)_\pi$ state at larger scattering angles $\theta^{\rm lab}_{d{\rm
He}} \ge 2^\circ$, while they are dominated by the $(1s)_\pi$ and
$(2s)_\pi$ states at forward angles. 
The peaks are well isolated and can be observed in the experiments with
the good energy resolution. 
Thus, we can conclude that we can obtain information of deeply bound
pionic 2$p$ state in addition to 1$s$ and 2$s$ states by observing the
spectra at finite angles. 
As indicated in Ref.~\cite{Ikeno}, the observation of several deeply
pionic bound states in a certain nucleus will help to deduce precise
information of pion properties and the chiral
dynamics at finite density~\cite{KSuzuki,Kolo,Jido}. 
We believe that our results provide a good evaluation for further experimental
studies of the states reported here, which should contribute to the
development of the field.

\section*{Acknowledgments}
We acknowledge the fruitful discussions with 
K. Itahashi, S. Itoh, T. Nishi and D. Jido. 
N. I. appreciates the support by the Grant-in-Aid for JSPS Fellows (No. 23$\cdot$2274).
This work was partly supported by the Grants-in-Aid for Scientific Research 
(No. 20540273 and No. 22105510).

%
%
%
%
%


\end{document}